\documentclass[a4paper]{jpconf}
\usepackage{graphicx}

\def\beq{\begin{equation}}
\def\eeq{\end{equation}}
\def\bey{\begin{eqnarray}}
\def\eey{\end{eqnarray}}

\def\lsim{\mathrel{\raise.3ex\hbox{$<$\kern-.75em\lower1ex\hbox{$\sim$}}}}
\def\gsim{\mathrel{\raise.3ex\hbox{$  $\kern-.75em\lower1ex\hbox{$\sim$}}}}

\def\grad{{\bf \nabla}}

\begin{document}
\title{An ecological approach to problems of Dark Energy, Dark Matter, MOND and Neutrinos }

\author{HongSheng Zhao}

\address{Scottish University Physics Alliance, University of St Andrews,  KY16 8SB, UK\\ Sterrewacht Leiden, P.O. Box 9513, 2300 RA Leiden, the Netherlands}

\ead{hz4@st-andrews.ac.uk}

\begin{abstract}
Modern astronomical data on galaxy and cosmological  scales have revealed powerfully the existence 
of certain dark sectors of fundamental physics, i.e., existence of particles and fields outside the standard models and inaccessible by current experiments.  
Various approaches are taken to modify/extend the standard models.  Generic theories
introduce multiple de-coupled fields A, B, C, each responsible for the effects of DM (cold supersymmetric particles), DE (Dark Energy) effect, and MG (Modified Gravity) effect respectively.  
Some theories use adopt vanilla combinations like AB, BC, or CA, and assume A, B, C belong to decoupled sectors of physics.  
MOND-like MG and Cold DM are often taken as antagnising frameworks, e.g. 
in the muddled debate around the Bullet Cluster.  
Here we argue that these ad hoc divisions of sectors miss important clues 
from the data.  The data actually suggest that the physics of all dark sectors is likely linked together by  a self-interacting oscillating field, which governs a chameleon-like dark fluid, 
appearing as DM, DE and MG in different settings.  It is timely to consider 
an interdisciplinary approach across all semantic boundaries of dark sectors, 
treating the dark stress as one identity, hence accounts for several ``coincidences" naturally.
\end{abstract}

\section{Fewer coincidences and dark sectors of the universe}

Dark Matter and Dark Energy are the most fascinating astronomical puzzles presented to modern physics of particles and gravitation.  To this date, the terms DM and DE are descriptive terms without a clear and unique underlying physics.  Finding a home for DM and DE in the edifice of symmetry-based physics is a challenge.  Although analogy is often drawn in the literature 
about DM, DE and Modified Gravity (MG) effects, the more fundamental links of these three effects have not been systematically shown.  
First many MG theories are DE in disguise where a special DE field is allowed to be non-uniform and non-minimally coupled to the metric.  The modification term enters the Einstein's equation
on the left or right as 
$G^{ab}-M^{ab} =  {8 \pi G \over c^4} T^{ab}_{known} , \rightarrow 
G^{ab} = {8 \pi G \over c^4} \left[ T^{ab}_{known} + N^{ab} \right],$
where  $T^{ab}_{known}$ is due to stress energy of known matter, 
e.g., neutrinos and baryons, $G^{ab}$ is due to the curvature tensor of the metric, 
and $M^{ab}$ is due to extra fields coupled to the metric,
and can equivalently viewed as the stress energy $N^{ab}= {c^4 \over 8\pi G } M^{ab}$ coming from New fields in the DE sectors \cite{LBMZ,Z06,HZL,Z081,ZL,Z082,ZF,Zmpd,Bek04,Sanders05,ZFS,S08,BM}.  

Second the vector, spinor or scalar in MG and DE 
are special kinds of charge-neutral DM fields where 
DM particles are allowed to self-interact in pairs (e.g., annihilate or be created from vacuum), 
but are forbidden to interact with fields of baryons.  This is because for MG and DE we have 
 $\nabla_b M^{ab}=0$ exactly for a Lagrangian construction for the DE or the coupling of MG field with the metric.  But $\nabla_b N^{ab} \approx 0$ for DM because of a very small rate of exchange of energy when DM and known matter collide.  So a DM field with a small cross-section to interact with itself and with the baryons is a more general description of MG and DE.  

Third MG, DE, DM fields can be described as a dark fluid with a varying equation of state in the framework of General Relativity.  
The proper treatment involves relativistic hydrodynamics rather than CDM-like 
N-body simulations because DM particles are not point masses moving on geodesics, 
they are packets of energy or fluid which propagates according to the equation of 
motion of the spin field, and the generally non-trivial coupling to itself and to the metric deflects
the path of DM particles from geodesics.  This deflection appears as a fluid-like pressure in the stress tensor.  Depending on the varying sound speed of the fluid and the Compton wavelength, 
the self-interaction can exhibit as short-range collisions by exchanging massive particles
or long-range fifth force by exchanging low-energy particle \cite{SS,FR,FP,KK1,KK2} or both
as in context of the lensing and velocity of the Bullet Clusters velocity \cite{ASZF,BCG,CBG}.
The fluid is a mixture of DE, Cold, Warm and Hot DM etc states, and 
needs not be stable, can make transitions among these as the sound speed changes with position
and redshift.  The energy of the dark fluid is nearly conserved, and its 
cosmic abundance of the dark fluid is nearly fixed at decoupling with 
the radiations.  In fact, a plausible candidate for the multi-phased dark fluid is the cosmic mixture of 
multi-flavored neutrinos and antineutrinos, which can make transition among at least several energy states, some very low in energy, some very high.  Already the uncertain low-energy physics of neutrino mass and flavor makes a plausible case for DE in the mass-varying neutrino theory (Mass-Varying Neutrinos, MaVaNs, by \cite{FNW,KNW,AZK,MPRW}.

Fourth the self-interaction cross-section must be small and a running function.  
Many theories introducing a zero or a fixed cross-section for self-interaction  
are excluded by data on either small scales (e.g., Cold DM) or large scales (e.g., Collisional DM).   Deviations from  the $\Lambda$CDM model must be small and 
gradual to preserve its success on Large Scale.
Previous multi-field models ignore the regularity of DM in galaxies and 
introduce more degrees of freedoms than justified by data. 
From the perspective of measuring weak lensing, dynamics or SNe distances,
astronomers detect only the bending of the space-time by the stress 
energy of all dark fields summed together, yet we see a cosmic ``coincidence" 
over $10^{30}$ order in length scale: everywhere a characteristic pressure exists 
$T_{11}\sim T_{22}\sim T_{33} \sim \Lambda c^4/G \sim (\Delta m_\nu^2)^2 c^5/\hbar^3$.  This pressure 
adapts to the environment satisfying a Tully-Fisher-Milgrom-McGaugh (TFMM) 
relation\cite{TFMM, M05a, M83}, where the cosmological constant $\Lambda$ is related  
to Milgrom's $a_0 \sim \Lambda^{1/2} \sim {\sigma_{\rm DM}^4 \over GM_{\rm baryon}}$.  
We expect this small pressure 
$T_{11}\sim(100$~km/s$)^2 \times10^9\,M_\odot/$kpc$^3$ 
can prevent the formation of Cold DM cusps in galaxies.  A possible home for 
MG and DE in fundamental physics might well be based on low-energy self-interactions of 
DM, especially what give neutrino mass difference $\Delta m_\nu^2 \sim(0.03$eV$)^2$.

For each galaxy, especially each spiral galaxy,
the radial distribution of DM mass $M_{DM}(R)$ and baryons $M_{bary}(R)$ 
could have enjoyed a large scatter as typically seen in CDM simulations depending on 
formation history, but instead the two  
strictly follow the very tight rule of Tully-Fisher-McGaugh globally and the formula of 
Milgrom \cite{TFMM,M05a,M83} at every observable radius without significant scatter.  
These rules hold tightly over many orders 
of magnitude scatter in the gas and stellar surface brightness and a wide range of 
formation history, including tidal dwarfs\cite{GFCKZT}.

On bigger scale 
the universe is in a state of very delicate balance, with an extremely fine-tuned ratio of the stress energy coming from the DM and DE, neutrinos and baryons.  These effects are all within about one order of magnitude of each other at the present instead of some 100 orders of magnitude 
apart if they come from truly independent physics.  The way to reduce fine-tuning is to insist 
symmetry or restrict degrees of freedom. E.g., 
these come from a common field coupling to all sectors with a cross-section comparable to that of neutrino self-coupling.  

\begin{figure}
\begin{center}
\includegraphics[width=12cm]{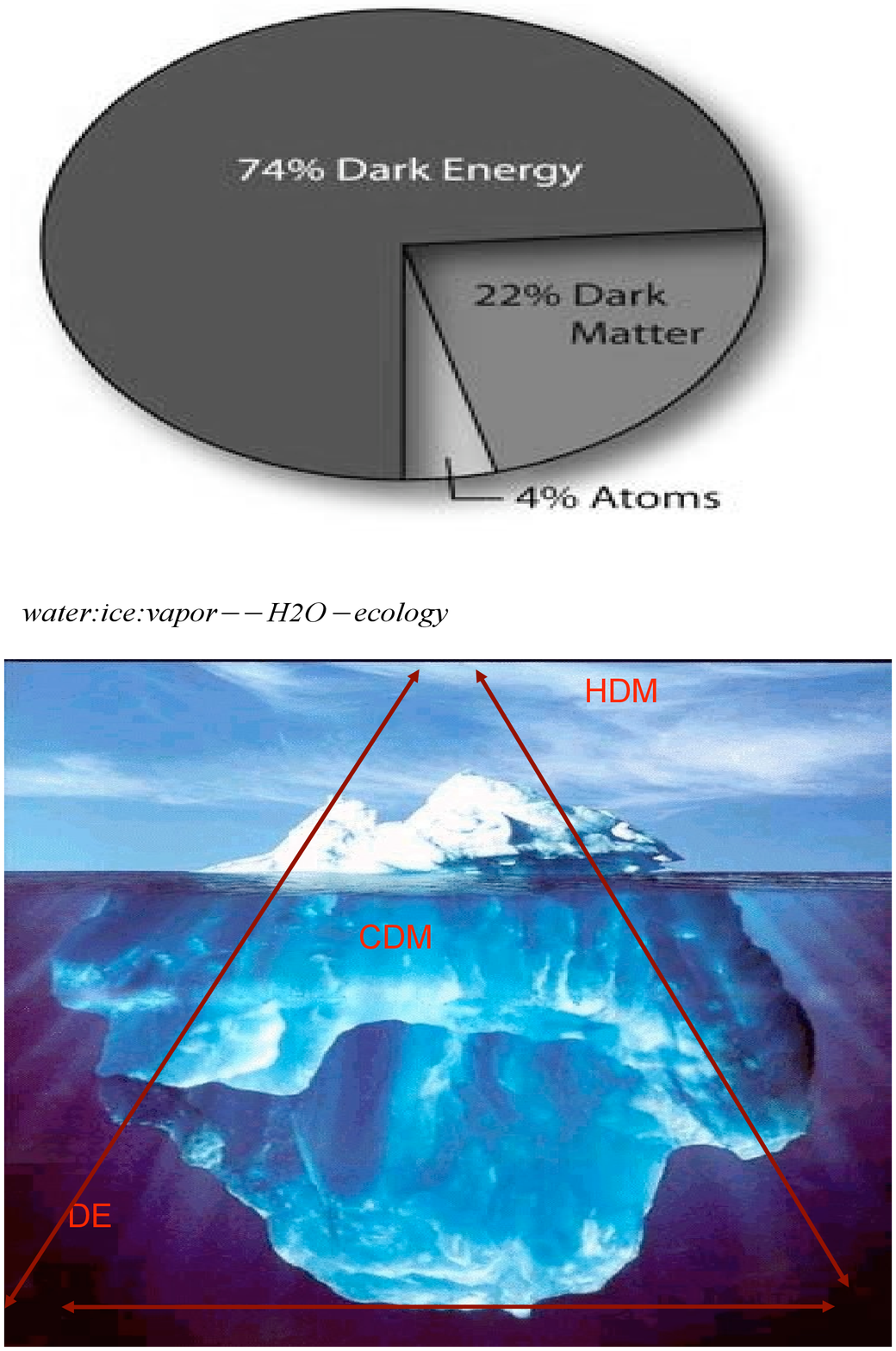}
\end{center}
\caption{\label{label} An ecological view of the Cosmos Most-wanted Pizza (CMP), where different dark components  (DE, Cold Dark Matter, Neutrino Hot Dark Matter) need not be separately conserved.  They might be phases of one Dark Fluid.  An analogy is drawn with the distribution of $H_2O$ molecule on earth in phases of water-ice-vapor; any theory with a built-in conservation of mass of ice would fail since ice melts.  A pure $\Lambda$CDM approach would fail to capture the possibility of non-conservation of (Cold) Dark Matter, and while an dynamical/ecological approach allows for the possibility of neutrino, CDM and DE to be phase change into each other, and explain the coincidence of scales in these three forms of non-baryonic energies in the universe.} 
\end{figure}

In short we argue for a theory-independent approach to DM, DE and MG. 
We propose to parametrize all these effects and the effects of neutrinos in terms of  
a Chameleon-like ``Dark Fluid'' ({\it CDF}) with an approximately conserved stress energy tensor $T_{\alpha\beta}$ plus GR.  It is a more systematic approach 
to DM interactions, since it combines the strength of previous methods using
purely MOND modified gravity, or Warm DM, or neutrino or Fifth-Force.  
A natural way to extend
$\Lambda$CDM is to build in the TFMM relation as the local minimum of
the action of the self-interacting DM fields.  If the DM-DE coupling
injects an uneven tiny pressure of order $\Lambda c^2/G \sim
10^9M_\odot/{\rm kpc}^3 \times (100{\rm km/s})^2$ to the random motion pressure of
Dark Matter in dwarf galaxies, this would suffice to remove the
undesired DM cusps in dwarf galaxies.  

The general inspirations of the models are illustrated by Figure.~1, where we seek to address the coincidence problems of DE, CDM/MOND and HDM (neutrinos) by one mechanism.       
In the following \S2 and \S3 we illustrate two possible approaches to unify the treatment of Dark Matter
and Modified Gravity.  \S4 illustrates a mathematical equivalence of these two treatments by a metric redefinition.

\section{Neutrino approach}

What gives mass to active neutrinos is an open question, and it is not even clear whether the mass is constant or not.  Interaction with Higgs-like field in other sectors or perhaps self-interactions in the neutrino sector (perhaps with sterile neutrinos) give an effective mass $m(z,x)$ to the neutrinos observed (chirially left-handed).  
This argues that the active neutrinos form a self-collisional fluid with free-streaming reduced by the pressure.
Since the exact mechanism of interactions are unknown, one could assume most generally that the effective mass could depend on environment.  

The original MaVaNs corresponds to 
a neutrino fluid of pressure 
$P = - f(n)$ so that the neutrinos can be modeled as a nearly polytropic fluid with an imaginary sound speed in an expanding universe.  
More generally the neutrino pressure $P$ is likely some function of a scalar field and  
its temporal and spatial gradient, $P = n E_0 \beta^2$, where $\beta=\beta(n, \dot{n}/n, |\nabla \Phi|)$ depends on the density $n$ and its rate of change and spatial gradient ($n^{-1} \nabla n$), and $E_0$ is  a constant energy scale. 

Unlike MaVaNs, here we assume that certain species of neutrinos have zero mass in the absence of perturbations, and the effects of mass appear only in collapsed systems, like galaxies and the solar system.   We assume that the cosmic fluid of neutrinos has a pressure $P$ due to interaction of   
a current vector $n u^\alpha$ of neutrinos
and a current vector $\bar{n} \bar{u}^\alpha$ of anti-neutrinos.  These two generally do not follow the same equations of motion due to their opposite spin-gravity coupling.  It is possible for 
these two currents to interact (creating a pressure) in a way reminiscent of how positive and negative charged particles interact in a plasma to keep average charge-neutrality on scales larger than the spacing of particles.  One expects $\bar{n} \sim n$, and a coupling like 
$\bar{u}^\alpha u^{\beta} R_{\alpha\beta}$ eventually reduces to terms like $\nabla_\alpha \bar{u}^\beta \nabla_\beta u^\alpha$, which in galaxies reduces to terms like $|\nabla \Phi|^2$.


To be specific we define $\beta$ to be the sound speed the relativistic neutrino fluid in units of the speed
of light $c$, and we adopt a variable sound speed such that  
\beq\label{pres}
\sqrt{n E_0 \over P} = {1 \over \beta}  = 1 + \sqrt{nE_0 \over P_g},
\quad 
P_g = {|\nabla \Phi|^2 \over 8\pi G}, ~{A^2 \over 8\pi G} \equiv  n E_0, 
\eeq
where $E_0$ is certain fixed low-energy threshold in the neutrino sector respectively, whose microphysics is not specified in detail, but could be thought as certain energy gap in neutrinos; note $\beta=0$ for the uniform universe with $\nabla \Phi=0$.
The quantity $1/\beta$ carries the meaning of the ratio of gravitational wave speed $c$ vs the sound speed $c \beta$ of the neutrino fluid, or the ``refraction index" of the neutrino fluid, which is assumed to be function of the ratio of the gravitational stress 
\beq
P_g = {|\nabla \Phi|^2 \over 8\pi G}
\eeq 
vs a characteristic stress $n E_0 \equiv {A^2 \over 8\pi G}$ of the neutrino fluid.  One can motivate such a correction by saying quantum effects must be considered 
to model interactions of the metric with a spin 1/2 neutrino fluid in a box of deBroglie wavelength  
$1/n \sim 1/n_0 \sim 5  {\rm mm}^3$; 
when the gravitational energy inside this box is just below or above certain threshold $E_0$, 
one might expect that anti-particles can disappear or appear (reminiscent of the famous Klein paradox), 
hence can change the  energy density of neutrinos and anti-neutrinos, hence the sound speed of the neutrino fluid.  

Interpreting in the picture of the the dielectric analogy of MOND \cite{BLT,FGBZ}
one would say that the neutrino medium is ``polarized" by a variable amount due to a varying gravitational stress.     
The adopted refraction index is such that the perturbations in the stress of the neutrino fluid propagate as the speed of light (as in vacuum) in strong gravity, but much slower in weak gravity;  no propagation or pressure in Minkowski space, where $\nabla \Phi=0$.  In the solar system, where $\nabla \Phi$ is much bigger than $A$, we have $\beta=1$.

We consider a classical Newtonian theory where one minimizes the action $S=\int dt L$ given by
\bey
S &=& \int dt \int dx^3  {\mathcal L},  ~{\mathcal L} =  \rho_b \Phi  + P_g -  P, 
\qquad P=(n E_0) \beta^2, 
\eey
where the relativistic pressure of neutrino fluid is included by a term $n E_0  \beta^2$, and 
$P_g + \rho_b \Phi$ 
is the Lagrangian of gravitational field and the baryons of density field $\rho_b$, which couples to gravitational potential $\Phi$.  Here we use the non-covariant formulation of gravity as an external field, which is the weak perturbation limit of General Relativity.  To be clear, 
$R$ is the scale factor, so that in the Lagrangian $L$ we are free to add or drop terms like $cst \int dx^3 R^3 n_b$ and $cst \int dx^3 R^3 n$, which are the total number of baryons or neutrinos respectively; the $cst$ here plays the role of the chemical potential.



The total Lagrangian now resembles the classical MOND of Bekenstein-Milgrom \cite{BM84} 
\beq
{\mathcal L} = {A^2 \over 8 \pi G} (y-F(y)) + \rho_b \Phi , 
\eeq
where 
\beq
F(y)  =  {y \over (\sqrt{y}+1)^2}, \qquad y  \equiv {|\nabla \Phi|^2  \over A^2}. 
\eeq

Apply the Lagrangian equation by varying the total Lagrangian with respect to $\Phi$, we obtain the  MOND-like Poisson equation
\beq
2 \nabla \left[ \mu \nabla \Phi \right] = 8 \pi G \rho_b,
\eeq
where
\beq
\mu = 1- dF/dy = 1- (1+\sqrt{y})^{-3} =1-(1+{ |\nabla \Phi| \over 3 a_0})^{-3}.
\eeq

It is interesting that a relativistic neutrino fluid with a non-trivial pressure can give the physics of MOND.
To match with MOND acceleration, we can set 
\beq
A =3 a_0 = 3.6 \times 10^{-10}{\rm m/s}^2.
\eeq
The resulting $\mu$ function here matches that of Zhao (2007), which are shown to be compatible with solar system data and spiral galaxy rotation curves, especially the Tully-Fisher-Milgrom relations. 
The MOND acceleration scale would corresponds to a mass or energy scale in the neutrino sector
\beq
E_0 = {3a_0^2 \over 8 \pi G n} \sim 2{\rm eV},
n= 200  {\rm cm}^{-3} .
\eeq  
Note that any eV range neutrinos with a normal non-relativistic pressure is  insignificant to galaxy potential.  The  non-linear coupling of the neutrino pressure to the metric through some unspecified quantum effects is the key.  In our picture neutrinos are not localized, this can create the effects of a ubiquitous dark matter fluid or modified gravity. 

\subsection{Relations of neutrinos, Higgs and vector field}

The above example belongs to a more general class of variable mass theory 
where one minimizes the action $S=\int dt L$ given by
\bey
S &=& \int dt \int dx^3  \sqrt{-g} {\mathcal L},  ~{\mathcal L} = \sum_f  n_f m_f  - M_p^2 R , 
\eey
where we use the natural units $\hbar=1$ and $c=1$, and 
$M_p^2 = {1 \over 8\pi G}$ is the Planck mass , $R$ is the Ricci scalar for the metric $g_{ab}$, 
and $n_f = \bar{\Psi} \Psi$ is the proper number density of the fermion of species $f$ and 
\beq
m_f \equiv \sqrt{ g_{ab} f^a f^b } 
\eeq
is the fermion's effective mass, which is determined by a vector field $f^a$ by 
\beq
f^a \equiv  s_f(\theta) m_0 u^a +   c_f(\theta) \dot{u}^a,\qquad \dot{u}^a \equiv u^b \nabla_b u^a, 
\eeq
where $\theta$ is an auxiliary field, and $u^a$ is a vector field of unit-norm in a metric $\tilde{g}_{ab}$, defined by 
\beq
\tilde{g}_{ab} u^a u^b =1, \qquad \tilde{g}_{ab} \equiv g_{ab} N^{-2}, \qquad N \equiv {h \over m_0}
\eeq
i.e., $p^a \equiv m_0 u^a$ resembles  the four-momentum of a particle of mass $m_0$ in 
the metric $\tilde{g}_{ab}$, which is related to the metric $g_{ab}$ by a conformal transformation factor $N^2 \gg 1$.  

To see the relation to the previous neutrino model, we can set  $N^2 = cst$ in present day galaxies, 
and set 
$s_f = cst = O(1)$, and $c_f=0$ for baryons so that  baryons have a fixed mass of order $N m_0$ and go on geodesics of $g_{ab}$, and 
so that $p^a = N m_0 g_{00}^{-1/2} (1,0,0,0)$ has a fixed norm in galaxies, where $g_{00}=1+2\Phi$ with $\Phi$ being the quasi-static gravitational potential.  Also set  
$s_f = {1 \over 3N}  (1-\varphi)^3 \ll O(1/N) $ and $c_f = \varphi \equiv \cos\theta$ for neutrinos so that neutrinos has a varying mass of order $m_0$.   We can minimize the action against the non-dynamical field $\theta$ to find 
\beq
2 (c_f^{-1}-2+ c_f)  = (N m_0)^{-2} \dot{u}^a\dot{u}^b g_{ab} 
\approx y \equiv \left({|\nabla \Phi| \over A} \right)^2, \qquad A \equiv m_0/N  
\eeq
in galaxies.  This resembles the MOND-like equation in \cite{ZL}.  So $c_f$ and the auxiliary field $\theta$ and the neutrino effective mass $m_f=m_0 \sqrt{F(y)}$ all track $y$ or the acceleration $|\nabla \Phi|$ with a characteristic scale $A=m_0/N$.    Minimizing against the metric $g_{00}$ or the potential $\Phi$, we find 
\beq
\nabla \cdot \left[ ( \mu \nabla \Phi \right] = 4\pi G \sum_f (n_f m_f), \qquad \mu=1- \alpha  {d\sqrt{F} \over dy}
\eeq
where $ \alpha \equiv {N^2 m_0 n_f \over M_p^2 m_0^2}={n_f m_f \over A^2 M_p^2}  $.  
These models have properties in between that of classical MOND, MaVaNs and Dark Matter.  

In Higgs-doublet models,  $h$ and $\theta$ are dynamical fields with a Lagrangian made of following 
kinetic and potential terms 
\beq
{\mathcal L}_h = \sum_f  (s_f h) n_f + g^{ab} \nabla_a h \nabla_b h  + h^2 g^{ab} \nabla_a \theta \nabla_b \theta + V(h)
\eeq
where a potential $V(h) \sim  cst (h^2 n_f m_0 M_p^{-2} - m_0^2)^2$ has a a minimum, can keep $\alpha=1$.   The Higgs field $h$ gives mass to baryons and neutrinos through the term $s_f(\theta)$, but in our case the symmetry of the Higgs doublet along the phase angle $0 \le \theta < 2\pi$ is broken 
by the term $c_f(\theta) \dot{u}^a$ spontaneously except in a Minkowski space where 
$\dot{u}^a=0$.  The scale $h = N m_0$ and the auxiliary field $\theta$ can take the meaning of the norm and the phase angle of a complex vector field $Z^a =  m_0 u^a \exp(i\theta)$ and $\bar{Z}^a = m_0 u^a \exp(-i\theta)$ with a (dynamical) norm $Z^a\bar{Z}^b g_{ab} = h^2$, as discussed in dark fluid models \cite{ZL}.    

\section{The vector approach: a Simple Lagrangian for MOND-like Dark Energy}

Here we illustrate how the roles of both DM and DE 
could be replaced by a vector field in a modified metric theory.  
This follows from merging two long lines of investigations 
pursued by Kostelecky, Jacobson, Lim and others on consequences of 
symmetry-breaking in string theory, 
and by Milgrom, Bekenstein, Sanders, Skordis and others 
driven by astronomical needs \cite{LBMZ,Z07,KS89}.

In Einstein's theory of gravity, 
the slightly bent metrics for a galaxy in 
an uniform expanding background set by the flat FRW cosmology is given by
\begin{equation}\label{metric}
g_{\mu\nu} dx^{\mu}dx^{\nu} 
=-(1+{2\Phi \over c^2}) d(ct)^2 + (1-{2\Psi \over c^2}) a(t)^2 dl^2 
\end{equation}
where $dl^2=\left(dx^2 +dy^2 + dz^2 \right)$
is the Euclidian distance in cartesian coordinates.
In the collapsed region of galaxies, the metric is quasi-static with 
the potential $\Phi(t,x,y,z)=\Psi(t,x,y,z)$ 
due to DM plus baryon, which all follow the geodesics of 
$g_{\mu\nu}$.   

Modified gravity theories are often inspired to preserve the Weak Equivalence Principle, 
i.e., particles or small objects still go on geodesics of above physical metric 
independent of their chemical composition.  Unlike in Einstein's theory, 
the Strong Equivalence Principle and CPT can be violated by, e.g., creating 
a preferred frame using a vector field by, e.g., a unit time-like vector field $U^{\mu}$  
which is designed to couple only to the metric but not matter directly.  It doesn't violate spatial rotation symmetry since it is time-like.  It has a kinetic Lagrangian with linear superposition of 
quadratic co-variant 
derivatives $\nabla (c^2U) \nabla (c^2U)$, where $c^2U^{\mu}$
is constrained to be a time-like four-momentum vector per unit mass by
$-g_{\mu\nu}U^{\mu} U^\nu= 1.$
The norm condition means the vector field introduces up to 3 new degrees of freedom; e.g., a perturbation in the FRW metric (Eq.\ref{metric}) has
$c^2U_{\mu} \equiv g_{\mu\nu}c^2 U^{\nu} \approx (c^2+\Phi,{A_x \over c},{A_y \over c},{A_z \over c})$, 
containing a four-vector made of an electric-like potential $\Phi$ and three new magnetic-like potentials.  But for spin-0 mode perturbations with a wavenumber vector ${\mathbf k}$, we can approximate 
$U_{\mu} - (1,{\mathbf 0}) \approx ({\Phi \over c^2},{{\mathbf k} V \over c})$, which contains 
just one degree of freedom, i.e., the flow potential $V(t,x,y,z)$.
We expect an initial fluctuation of $c|{\mathbf k}|V \sim |\Phi| \sim c^2 N^{-1} \equiv 10^{-5}c^2$ 
can be sourced by a standard inflaton; the vector field tracks the spectrum of metric perturbation.

When it comes to writing down a specific Lagrangian of the vector field,  simplicity is the guide since GR plus simple $\Lambda$CDM largely works.  
Let's start with forming two pressure terms for any four-momentum-like
field $A^{\mu}$ with a positive norm 
$m c^2 \equiv \sqrt{-g_{\alpha\beta}A^{\alpha}A^{\beta}}$ by 
\begin{equation}\label{Z}
8\pi G {\cal J}(A) \equiv 
{1 \over 3} \left({\nabla_\alpha A^\alpha \over m}\right)^2,~
8\pi G {\cal K}(A) \equiv {\nabla_\parallel A^\alpha \over m} {\nabla_\parallel A_\alpha \over m}
\end{equation}
where the RHSs are co-variant with dimension of acceleration squared,
and $\nabla_\parallel=A^\alpha \nabla_\alpha$ or $\nabla_\alpha$ stands for the co-variant derivative with space-time coordinates 
along the direction of the vector $A$ or the dummy index $\alpha$ respectively.  
From these we can generate two simpler 
pressure terms $K$ and $J$ of the unit vector field $U^\alpha$ by
\begin{equation}\label{KJgalaxy}
\begin{array}{cllcll}
J \equiv {\cal J}(U) &\sim&0,                 & K \equiv {\cal K}(U) &\sim&{|\nabla \Phi|^2 \over 8\pi G}  ~\mbox{\rm in galaxies}\\
 &\sim&{3 c^2H^2 \over 8\pi G},               &   &\sim&0  ~\mbox{\rm in flat universe}
\end{array}\label{KJcosmo}
\end{equation}
where the approximations hold for $U^\alpha$ with negligible 
spatial components and nearly flat metric (Eq.\ref{metric}).
Note the $J$ and $K$ are constructed so that we can control 
time-like Hubble expansion and space-like galaxy dynamics {\it separately}.  The $K$-term, with a characteristic pressure scale ${a_0^2 \over 8\pi G} =P_0$ in galaxies, is the key for our model.  The $J$-term, meaning critical density, has a characteristic scale $N^2P_0 \sim 10^{10} P_0$:
at the epoch of recombination $z=1000$ when baryons, neutrinos, and photons contribute 
$\sim (8,3,5) \times 10^9 P_0$ respectively to the term $J={3c^2 H^2 \over 8 \pi G}$; 
so the epochs of equality and recombination nearly coincide.

Now we are ready to construct our total action $S=\int d^4x |-g|^{1\over 2} {\cal L}$ 
in physical coordinates, where the Lagrangian density
\begin{equation}\label{action}
{\cal L} = {R \over 16 \pi G} + L_m + L_J + L_K + (U_\nu U^\nu \!\! +1) L^m, ~
\end{equation}
where $R$ is the Ricci scalar, $L_m$ is the ordinary matter Lagrangian.
For the vector field part, $L^m$ is the Lagrangian multiplier for the unit norm and
we propose the new Lagrangian 
\begin{equation}\label{LJ}\label{LK}
L_J = \int_{0}^{J} \!\!    d J  \lambda_\infty(x)|_{x=\sqrt{|J| \over P_0}} , \qquad
L_K = \int_{\infty}^{K} \!\!  d K \lambda(x)|_{x=\sqrt{|K| \over P_0}}, 
\end{equation}
where the non-negative continuous functions 
$\lambda(x) =\left(1+{x \over 3} \right)^{-3} -0$, and $\lambda_\infty= 1-(1+\infty/\infty)^{-3}=1-\mu_B$, where  $\mu_B \equiv 2^{-3}=1/8$.   A more fine-tuned parametrization is given in Zhao (2007), which passes the BBN constraints better.  

Taking variations of the action with respect to the metric and the vector field, 
we can derive the modified Einstein's equation (EE) and the dynamical equation for the vector field.  The expressions are generally tedious, 
but the results simplifies in the perturbation and matter-dominated regime that interest us. 
As anticipated in \cite{Lim04}  the $ij$-cross-term of EE yields $\Psi-\Phi=0$ for all our models, which means
incidentally twice as much deflection for light rays as in Newtonian.
The tt-equation of Einstein reduces to the simple form
\begin{eqnarray}\label{poisson}
4 \pi G \rho &=& \grad^2\Phi - \grad \cdot \left[\lambda_n\left({|\grad \Phi|\over a_0}\right)\grad \Phi\right],~\mbox{\rm in galaxies} \\
{8 \pi G \bar{\rho} \over 3\mu_B} &=& H^2 - {\Lambda_0 \over 3\mu_B}, ~\mbox{\rm in matter-dominated FRW}
\label{hubble}
\end{eqnarray} 
Here the pressure from the vector field creates new sources for the curvature.
The term ${\grad(\lambda_n(x)\grad \Phi) \over 4\pi G}$ in the Poisson equation acts as if adding DM for quasi-static galaxies.  A cosmological constant in the Hubble equation is created by 
\begin{equation}\label{cosmocst}
{\Lambda_0 c^2 \over 8\pi G} = -\!\!\int_{\infty}^{0} \!\!\! \lambda(x) d(P_0x^2) \approx {(3P_0)^2 } 
\end{equation}
 
{\it For binary stars and the solar system},  
$4 \pi G \rho  -  \grad^2\Phi \approx 0$ is true 
because the gravity at distances 0.3AU to 30AU from a Sun-like star is much greater than the maximum vector field gradient strength $a_0$, so ${dL_K \over dK}=0$; in fact, $|\grad \Phi| \approx {G M_\odot \over r^2} \sim (10^9-10^5) a_0$, and the typical anomalous acceleration is $\sim 10^{-10}a_0$, well-below the current detection limit of $10^{-4}a_0$  (Soreno \& Jezter 2006).  This might explain why most tests of non-GR effects around binary pulsars, black holes and in the solar system yield negative results;  Pluto at 40 AU and the Pioneer satellites  at 100 AU might show interesting effects. Extrapolating the analysis of \cite{FJ}, we expect GR-like PPN parameters and gravitational wave speeds in the inner solar system.

{\it Near the edges of galaxies}, we recover 
the non-relativistic theory of Bekenstein \& Milgrom \cite{BM84} with a function  
\beq\label{mu}
\mu(x)  \equiv  1-\lambda_n(x)
   \sim \mu_{min} + x, ~\mbox{\rm if $x={|\grad \Phi| \over a_0} \ll 1$}.
\eeq
Note that  $\mu(x) \rightarrow x$ hence rotation curves are asymptotically flat except for a negligible correction $\mu_{min} \sim 10^{-15}$.  
In the intermediate regime $x=1$ our function with $1-\lambda_n(x) \sim (0.55-0.6)$ for $n=2-5$ respectively.  Galaxy rotation curves prefer a relatively sharper transition than $\mu(x)=x/(1+x)=0.5$ at $x=1$ \cite{FGBZ}
where we can identify $g_B/(g_{DM}+g_B) = \mu(x)$.  So our model should fit observed rotation curves.

{\it For the Hubble expansion:} the vector field creates cosmological constant-like term  
${\Lambda_0 c^2\over 8\pi G} \approx 9 P_0 $ below the zero-point of the energy density 
in the solar system because 
the zero point of our Lagrangian (Eq.\ref{LK}) is chosen at $N^2 P_0 \le K <+\infty$.
During matter domination,  the contribution of matter $8 \pi G \rho$ and $\Lambda_0$ to the Hubble expansion $H^2$ (Eq.\ref{hubble}) is further scaled-up 
because the effective Gravitational Constant $G_{eff}=G/\mu_B = 8 G \ge G$.
Coming back to the original issue of the $3:1$ ratio 
of matter density to our cosmological constant, Eq.(\ref{hubble}) predicts that 
${\Lambda_0 c^2 \over 8\pi G \mu_B}: {\bar{\rho}_b c^2 \over \mu_B} 
\sim {9 P_0 \over \mu_B}: {4 (1+z)^3 P_0 \over \mu_B}$, which is close to the desired $3:(1+z)^3$ ratio.  
Adding neutrinos makes the explanation slightly poorer. 
So the DE scale is traced back to a separate coincidence of scale, 
i.e., the present 
baryon energy density $\bar{\rho}_b c^2 \sim 4 P_0$, where $P_0$
contains a scale $a_0$ for the anomalous accelerations on galactic scale.
This model predicts that DE is due to a constant of vacuum.

In our model, the effective DM (the dog) follows the baryons (the tail) throughout the universal $(1+z)^3$ expansion with a ratio set by the parameter $\mu_B$.  
To fit the $\Lambda$CDM-like expansion exactly, we note
the Hubble equation for a flat FRW cosmology with vector field
and standard mix of baryons, neutrinos and photons 
${\Omega_b h^2 \over 0.02} \approx {\Omega_{\nu}h^2 \over 0.002}{0.07{\rm eV} \over m_\nu} \approx {\Omega_{ph} h^2 \over 0.000025} \sim 1$ yields at the present epoch
\beq\label{Omega}
{\Omega_b + \Omega_\nu + \Omega_{\rm ph} \over \mu_B} 
= 1-{\Lambda_0 \over 3\mu_B H_0^2} =  \Omega_{m}^{\Lambda CDM}
\eeq
The 2nd equality fixes $\mu_B^{-1}=(8-8.4)$ 
if we adopt $a_0/c \approx H_0/6 \approx 12$km/s/Mpc and 
$\Omega_{m}^{\Lambda CDM}=(0.25-0.3)$.
The 1st equality would predict an uncertain but very small neutrino mass 
$m_{\nu} \sim \pm 0.3$eV, consistent with zero. 
So the role of neutrinos in uniform expansion can be completely replaced by the vector field.

\section{A unified framework for Dark Matter, Dark Energy and 
Modified Gravity}

Finally we illustrate the relation between Modified Gravity and Interacting Dark Matter.

\subsection{Einsteinian gravity with an interacting Dark Matter field}

Let's consider Einsteinian gravity but with the normal matter 
(of standard model of particle physics) 
being coupled to the field of a dark matter particle.
Let the dark particle be spin-1, hence it is described by a vector field, hence with 4 degrees of freedom.
Unlike the spin-1 photon field, which is a massless gauge-invariant vector field with zero expectation value in vacuum, the dark particle vector field is given a unit norm, hence it is a massive field and has a non-zero expectation value in vacuum.  
The vector field is given self-coupling and coupling to normal matter, which break the gauge-invariance.

The gravity sector is now simply described by a metric $g_{ab}$ with a sign convention (+,-,-,-) and its associated Ricci scalar $R$, hence the Einstein-Hilbert action plus matter,
\beq
S  = S_g + S_m = -\int dx^4 \sqrt{-g} \left[ {R \over 16 \pi G} + L_m \right],\qquad
L_m= L_A + L_{int} + L_J 
\eeq
where matter is consisted of the
Lagrangian density $L_J$ for a pressure-less 
matter fluid with matter flux vector $J^a = \rho u^a$, 
and $L_A$ for the dark matter vector field $A^a$, and
an interaction term of the two vector fields.  
Specifically
\bey 
L_J & =& \sqrt{ g_{ab} J^a J^b } - \phi \nabla_a J^a ,
\eey
where $\phi$ is a Lagrangian multiplier field for the conservation of matter flux 
$J^a = \rho u^a$ of a collisionless dust of density $\rho$ and four-velocity $u^a$.
It interacts with the vector field $A^a$ via  
\bey
L_{int} & = &  C^2 \sqrt{BC}
\sqrt{ g_{ab} J^a J^b g_{cd} A^cA^d - (1 - B) (g_{ab} J^a A^b)^2} - \sqrt{ g_{ab} J^a J^b } 
\eey
where $B$ and $C$ are coupling constants.  The vector field contributes via 
\beq
L_A = { m^2 \over 2} F_{ab}F^{ab}  + (1- g_{ab} A^aA^b ) \lambda 
\qquad F_{ab} = g_{bc} \nabla_a A^c  - g_{ac} \nabla_b A^c, 
\eeq
which consists of a kinetic photon-like term from Faraday tensor $F_{ab}$, and a massive potential in the form of a unit norm constraint for $A^a$ and a Lagrangian multiplier $\lambda$.
And to make the argument simpler we assume the mass $m$ and 
$\lambda \propto m^4$ is very small, $\beta = 16 \pi G m^2 \ll 1$,  
so that we can neglect the dark matter term $L_A$, keeping only the interaction term $L_{int}$.  

Note that our Lagrangian $L_J$ depends on $g_{ab}$ in a non-linear fashion.  However,
non-linearity is not a sufficient condition for modified gravity.  It can be 
shown that the stress tensor associated with $L_J$ is 
$T_{ab} = {\delta (L_J \sqrt{g}) \over \sqrt{g} \delta g^{ab}} = J_a J_b (g^{ab}J_aJ_b)^{1/2} = \rho u^a u^b$, as expected for a collisionless dust.

Note that we cannot observe the dark matter $A^a$ field directly, it is observed through 
its interaction with baryonic dust $J^a$.  Here we considered only a species of dust.
If we generalize for the interaction/coupling to be the same for all species
of the baryonic dust, then we cannot detect the dark matter 
through differential measurements of the baryons (the strong equivalence principle).
However, the dark matter needs not track the baryonic dust exactly, 
so we don't expect the baryonic mass center to coincide with its kinematic (gravitational) center (Kesden \& Kamionkowski 2006).
 
\subsection{``Modified" gravity in redefined metric}

Alternatively one can redefine the metric 
\bey
\tilde{g}_{ab} &=& \left( g_{ab} - (1 - B) A_a A_b  \right) C \\
\tilde{g}^{ab} &=& \left( g^{ab} - (1- B^{-1}) A^a A^b  \right)/C.
\eey
The nice feature is that the matter action is now simplified to that of a 
pure matter field $J^a$, with a new Lagrangian density 
\beq
\tilde{L}_J = \left[\sqrt{\tilde{g}_{ab} J^aJ^b} - \phi \nabla_a J^a \right],
\eeq
which is completely decoupled from the vector field $A^a$.  
However, the new Ricci scalar, formed out of second derivatives of the new metric, 
differs from the old Ricci scalar by a K-term, so that the gravity is ``modified"
with an effective gravitational constant $\tilde{G}$ given by 
\beq
{\tilde{G} \over G} = {\sqrt{-\tilde{g}} \over C \sqrt{-g}} = \sqrt{B} C.
\eeq   
Rewriting the action S  in the new metric, we find an action  
resembling that of a ``modified" gravity, specifically 
\beq
S = \tilde{S}_g + \tilde{S}_J =  -\int dx^4 \sqrt{-\tilde{g}} \left[ 
 {\tilde{R} + \tilde{K} \over 16\pi \tilde{G} }\right] - \int dx^4 \sqrt{-\tilde{g}} \tilde{L}_J, 
\eeq
where 
\beq
\tilde{K} = \left[ K^{ab}_{mn} \nabla_a A'^m \nabla_b A'^n \right] + (\tilde{g}_{ab}A'^aA'^b-1) \lambda ,
\eeq
\beq
A'^a \equiv A^a/\sqrt{BC}
\eeq
\beq
 K^{ac}_{bd} \equiv (c'_1 \tilde{g}_{ab} + c'_4 A'^aA'^b) \tilde{g}_{mn} 
+ (c'_2 \delta_{m}^{a} \delta_{n}^{b} + c'_3 \delta_{n}^{a}\delta_{m}^{b}) 
\eeq
where $\lambda$ is a Lagrangian multiplier field,
\beq
c'_1 = -c'_4 = -{1-B^{-1} \over 2} c'_2 + B \beta, \qquad -c'_3- c'_1 = c'_2 = B-1.
\eeq
where we have considered 
more generally when the mass $m$ of the vector field is not small, the c-parameters are functions of $\beta =16 \pi G m^2$ as well \cite{Foster05}.

The redefined action is that of a special case of the Einstein-Aether modified
gravity, where there is no interaction between normal matter and the vector field (called aether).  Assuming $A^a$, $J^a$ and $\tilde{g}^{ab}$ as independent freedoms as conventionally done, the Einstein equations are obtained; taking its trace we have
\beq
2 \tilde{R} -  \tilde{g}^{ab}T^A_{ab}  = 8 \pi \tilde{G} \tilde{T}^J, \qquad
\tilde{T}^J = {\tilde{g}^{ab} J_a J_b \over \sqrt{\tilde{g}_{ab} J^a J^b} } 
\eeq
where $T^J$ is the trace of the stress tensor of the collisionless dust,
and $\tilde{R}$ is the Ricci scalar, 
$-T^A_{ab}$ is the part of Einstein tensor involving a fairly lengthy expression of 
second derivatives of the vector field $A^a$ and the metric $\tilde{g}$.
For the uniform expansion of the universe in co-moving coordinates $(t,x,y,z)$, 
the metric is given by 
\beq \tilde{g}_{ab} dx^a dx^b =  dt^2 - a(t)^2 (dx^2+dy^2+dz^2),\eeq
the matter current $J^a = (\rho, 0, 0, 0)$ 
and $A'^a=(1,0,0,0)$ with a preferred time-like direction.
The Hubble equation is unchanged, 
\beq
3 \left( {d a(t) \over a(t) dt} \right)^2 = 8 \pi {G \rho \over \mu_C}, \qquad
{1\over \tilde{\mu}_C} =  {\tilde{G}/G \over 1+ (c_1+3c_2+c_3)/2} ={C \over \sqrt{B} },
\eeq
except for a correction $\mu_C$ of the effective gravitational constant.
For weakly perturbed metric near a static galaxy or the solar system  with 
\beq \tilde{g}_{ab} dx^a dx^b = (1+2\Phi) dt^2 - (1-2\Phi) (dx^2+dy^2+dz^2),  \eeq
the matter current $J^a=(\rho,0,0,0)$, and the vector field $A'^a=(1/\sqrt{\tilde{g}_{00}},0,0,0)$ with a preferred time-like direction, and
the Poisson equation is changed to 
\beq
\nabla^2 (2 \mu_N \Phi) = 8 \pi G \rho, \qquad
{1 \over \mu_N} =  {\tilde{G}/G \over 1-(c_1+c_4)/2 } = {C \sqrt{B} }
\eeq
with a factor $\mu_N$ correction of the effective gravitational constant.

One can create appropriate amount of dark matter-like effects by 
selecting the factors $\mu_N$ and $\mu_C$ in solar system, 
in galaxies and in matter-dominated cosmology.  
The above arguments can be generalized to the case where $\mu_N$ and $\mu_C$ are 
scalar fields, which varies with redshift and position
to resemble the $V-\Lambda$ model and 
$F(K)$ models of co-variant MOND \cite{HZL} and various 
modified gravities.      
E.g., the modified source gravity of Carroll et al. can be recovered by
models with $\mu_C=\mu_B = C $ being non-dynamical scalar fields tracking the Ricci scalar through an added potential term $V(C)$ in the interaction Lagrangian.

In short, the above example  suggests that dark matter and modified gravity can be 
the two view points of the same phenomena.  Our vector field can be 
viewed as spin-1 dark matter field in Einsteinian gravity and $g_{ab}$, and this  
dark matter field $A^a$ is co-variantly 
coupled to the luminous matter current field $J^a$.  However, this 
vector field is not exactly the cold dark matter, and in fact it does not condense 
in galaxies in our model, but it does decelerates the expansion of the universe, and  
contributes to the critical density.  
Alternatively, one can view our vector field as a field modifying the gravity sector
of metric $\tilde{g}$.  
In this case, the luminous matter is decoupled from the vector field $A^a$.

In short the coincidences of scales of DM, DE and Neutrinos are intriguing.  We advocate that it is theoretically satisfying to find a unified solution to these problems at a fundamental level.  

\section{Acknowledgments}
The author  acknowledges discussions with Baojiu Li, Benoit Famaey, Bob Sanders, Eugene Lim, 
A. Kostelecky, Sean Carroll, Subir Sarkar and David Spergel.   Thanks also to the simulating Monday pub with Keith Horne, Andrew Green and Martin Feix, and to many hospitalities at  Observatoire Astronomique de Strasbourg,  Institut d'Astrophysique de Paris and the Dark Cosmology Center of Copenhagen University.


\end{document}